\documentclass[%
 reprint,
 amsmath,amssymb,
 aps,
]{revtex4-1}

\usepackage{xcolor}
\usepackage{soul}
\usepackage{graphicx}
\usepackage{dcolumn}
\usepackage{bm}

\usepackage{capt-of}

\begin{document}

\preprint{ver-20200118}

\title{Twisted double-layer chiral spin structures in anti-ferromagnetic delafossite PdCrO$_2$}

\author{Changhwi Park}
\author{Jaejun Yu}%
\affiliation{%
Center for Theoretical Physics, Department of Physics and Astronomy, Seoul National University, Seoul 08826, Korea\\
}%

\date{\today}

\begin{abstract}
We predict that the anti-ferromagnetic (AFM) magnetic structure of delafossite PdCrO$_2$ has a staggered chirality with the double-layer periodicity along the $z$-axis with intriguing magnetic anisotropy. We study the electronic and magnetic structures of PdCrO$_2$ by carrying out density functional theory (DFT) calculation. The calculated ground state turns out to exhibit a 120$^\circ$ AFM ordering with staggered chirality. The result is consistent with the previous neutron experiments, but our calculations determine and provide more detailed information on the high symmetric easy-plane and local-easy-axis structure. We investigate the origin of staggered chirality as well as the easy-plane and local-easy-axis directions. Based on the DFT and effective spin interaction models, we demonstrate the complex magnetic ordering requires novel four-spin interactions, which cannot be described by the usual exchange interaction. Also, we show Pd electronic bands near the Fermi level pose a tiny degeneracy breaking, which arises from the magnetic anisotropy with different easy-plane and local-easy-axis. We expect that the twisting and rotating modes of easy-planes in the magnetic double layers can be the critical factor to understand a hidden magnetic property of PdCrO$_2$ and its related macroscopic property like the previously reported anomalous Hall effect.
\end{abstract}

\maketitle


\section{\label{sec:level1} introduction \protect\\ }

The metallic delafossite PdCrO$_2$ is known to be a two-dimensional triangular Heisenberg anti-ferromagnet (AFM) \citep{2009_Takatsu_PdCrO2,2010_Takatsu_PdCrO2,2013_JongMokOk_PdCrO2,2014_Takatsu_PdCrO2,2014_HanJinNoh_PdCrO2,2015_DavidBillington_PdCrO2,2018_DucLe_PdCrO2,lechermann2018hidden,2013_Sobota_PdCrO2} with $T_N=37.5$ K. It has alternating layers of Pd and CrO$_2$ triangular lattices, respectively. The triangular lattice of Pd atoms construct metallic layers, responsible for the high conductivity within the layer along the $xy$-directions,
while the CrO$_2$ layers are expected to be insulating layers with the local magnetic moments of Cr$^{3+}$($S$=3/2). The reported a neutron diffraction peaks at $(hkl)=(1/3, 1/3, n/2)$ \citep{2009_Takatsu_PdCrO2} indicate that PdCrO$_2$ has a $\sqrt3 \times \sqrt3 \times 2$ supercell magnetic structure, e.g., AFM unit-cell consisting of three Cr atoms within the layer and the multiple of double-layer structures along the $z$-direction. This magnetic supercell with the 120$^\circ$-ordering of three Cr local moments is consistent with Yamaji angles \citep{2017s_ghannadzadeh_PdCrO2}, quantum oscillations \citep{2013_JongMokOk_PdCrO2}, and a direct observation by angle resolved photoemission spectroscopy (ARPES).

Although the $120^\circ$ magnetic ordering of Cr$^{3+}$($S$=3/2) moments is expected to be the basic structure,
there are still a lot of controversies on its details.
Takatsu and coworkers \citep{2014_Takatsu_PdCrO2} suggested a series of non-collinear 6-layer magnetic structures.
Some of them are based on the previous study of LiCrO$_2$ \citep{1995_kadowaki_LiCrO2}, but others have more intricate details.
A spin scalar chirality model, which contains non-coplanar spin ordering, was suggested as a possible candidate.
However, it is still under controversy since direct evidence of scalar chirality is not observed. Further, scalar chirality is not the only origin of anomalous Hall effect \citep{suzuki2017cluster,nakatsuji2015large}.

Among the Takatsu's 6-layer in-planar models, a staggered chirality model is a strong candidate.
Their rotation direction can define the chirality of three in-planar local magnetic moments.
In the system with three in-planar spin moments with 120$^\circ$ AFM ordering, there are only two rotation directions, i.e., handedness or chirality.
The chirality can be defined by its stacking ordering of assigned atom indices and the rotation direction of spins of them. Further, depending on the stacking sequence of chiralities, the system can have multi-layer periodicity along the $z$-axis. Indeed, Duc Le and coworkers \citep{2018_DucLe_PdCrO2} suggested a simpler model, a staggered chirality with two layers. Not only the 2-layer model reduces the number of model parameters, but it also gives better fitting quality for neutron inelastic scatterings and linear spin-wave theory \citep{2015_Toth_linear}.
By employing the linear spin-wave theory with seven fitting parameters, they have successfully suggested a Heisenberg model. However, although their interpretation describes well neutron inelastic scattering data, that model still has a limitation. The Heisenberg interactions only cannot distinguish the energy difference of staggered and straight chiralities. Also, there is another remaining issue related to the magnetic anisotropy-energy related to easy-plane and local-easy-axis directions. Indeed, the easy-plane and easy-axis direction of the ground-state magnetic configuration is still ambiguous. Further, there is no apparent clue for easy-plane and local-easy-axis problems.

The Fermi surface of PdCrO$_2$ has a nearly hexagonal shape \citep{2014_HanJinNoh_PdCrO2}.
Similar $AB$O$_2$ delafossite compounds, which have Pd or Pt atoms on $A$-site, also have a hexagonal shape of Fermi surfaces \citep{2017_mackenzie_delafossite}, yet PtCoO$_2$ has a concave shape compared to PdCoO$_2$ \citep{2009_KyooKim_PdCoO2,2009_HanJinNoh_PdCoO2}.
The $xy$-planar resistivity of PdCrO$_2$ at room temperature (295K) was found to be very low, $\approx 9 \mu\Omega$ cm, while its $z$-directional resistivity is high ($\rho_c/\rho_{ab} \ge 150$) \citep{2015_Hicks_PdCrO2,2010_Takatsu_anisotropy}.
In contrast with PdCoO$_2$, which does not have magnetic moments on $B$-site, an intriguing feature of PdCrO$_2$ is its band folding, magnetic ordering, and unconventional anomalous Hall effect \citep{2010_Takatsu_PdCrO2}.
There is also an issue on the folded Fermi surface.
The spectral weight of the folded Fermi surface is rather weak, so there should be an explanation for it.
A recent theoretical study suggested that the band-folding effect of PdCrO$_2$ originates from Kondo lattice Hamiltonian \citep{sunko2018probing}.
They explained that the weakened spectral weight of folded bands is caused by strong Coulomb repulsion $U$ on Cr atoms.

Here, we investigate the electronic and magnetic structures of PdCrO$_2$ by carrying out non-collinear-spin density-functional-theory (DFT) calculations.
We also set up a tight-binding Hamiltonian to describe the
electronic bands and Fermi surfaces, which depend on its magnetic structures.
We suggest an effective spin model for the observed and calculated magnetic structure.
We also demonstrate that the magnetic structures of Cr atoms affect both the $z$-directional electron hopping and the tiny degeneracy breaking on Pd bands.

\begin{figure}
\includegraphics[width=0.3\textwidth]{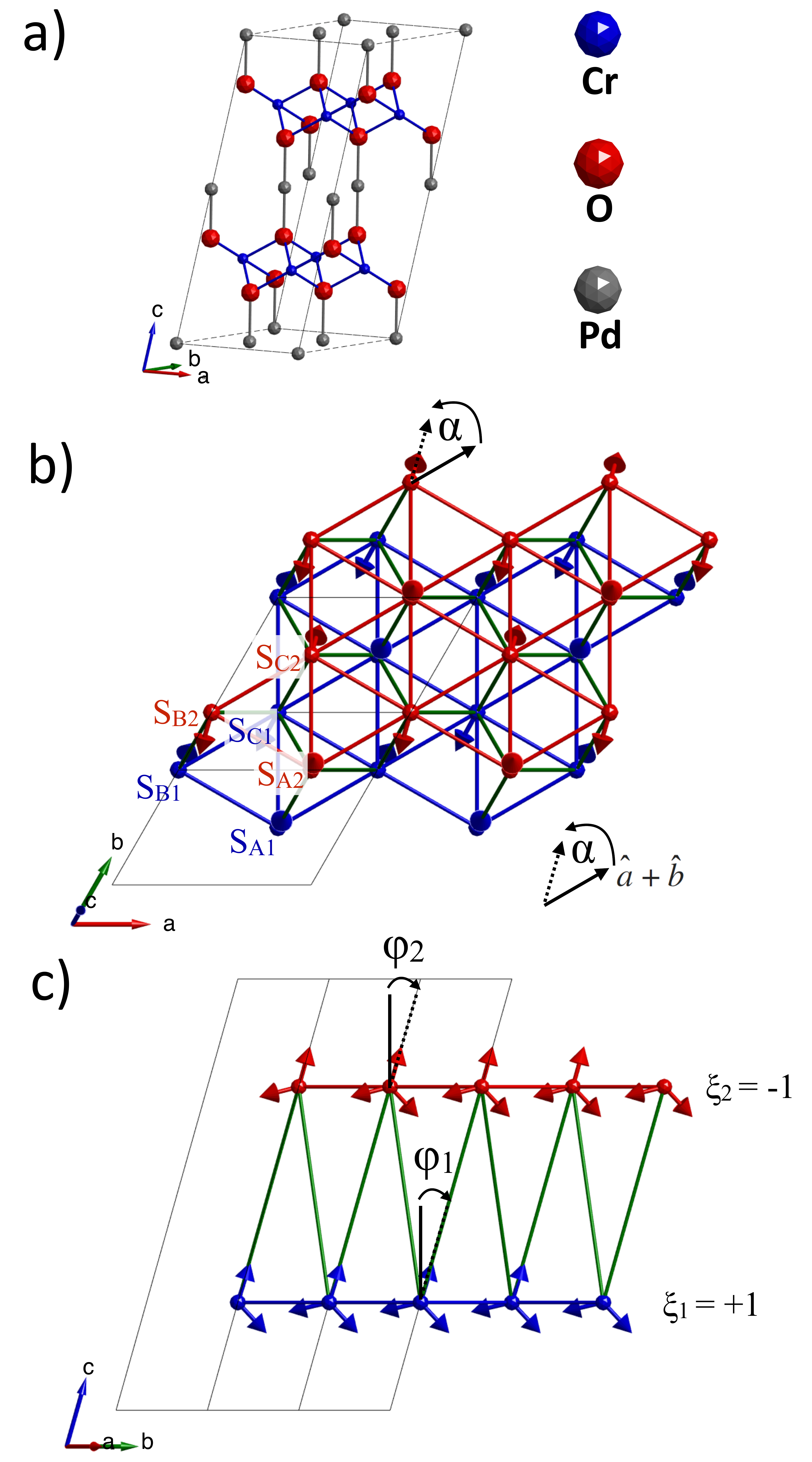}
\caption{\label{fig:structure}
Unit cell and a magnetic structure of PdCrO$_2$.
a) The 2-layers cell with 3-Cr atoms in one layer is the smallest unit cell to represent a staggered chirality.
b),c) An example magnetic structure of $\alpha_1 = 31$, $\alpha_2 = 44$, $\phi_1  = 17$, $\phi_2 = 16$, $\xi_1= +1$, $\xi_2=-1$.
The 1st layer is painted blue, 2nd layer is painted red.
Arrows indicate spin moments of Cr$^{3+} $.
1st nearest inter layer interactions are colored in green.
2 layers have their easy-planes ($\alpha$) and easy-axes ($\phi$) independently, which are illustrated in Eq. (\ref{eq:equation 1}).
When $\alpha = 30^\circ,150^\circ,270^\circ$, the spin easy-planes align to the direction of 1st nearest inter layer interaction connections.
}
\end{figure}

\section{methods}

We carried out non-collinear-spin DFT calculations to determine a series of magnetic configurations and their corresponding total energies.
We used the OpenMX code \citep{2003_Ozaki_variationally,2004_Ozaki_numerical,2006_han_n}.
The $\sqrt3 \times \sqrt3 \times 2$ supercell was used to describe its magnetic configurations (Fig.~\ref{fig:structure}). Minimum $32 \times 32 \times 90$ Ngrids and $10 \times 10 \times 10$ Kpoints grids are used for fast calculations.
Maximum $64 \times 64 \times 180$ Ngrids 
and $14 \times 14 \times 14$ Kpoints
are used for the convergence checking.
We used $s^2p^2d^2f^1$ pseudo-atomic orbitals for Pd and Cr atoms, and $s^2p^2d^1$ for O atoms.
The SCF-criterion of $4 \times 10^{-8}$ Hartree / 6-Cr atoms is used to guarantee the energy convergence for the spin configuration of easy-plane and local-axis rotation modes (approximately 10$\mu$eV / 1-Cr atom).
A penalty-function-constraint method \citep{kurz2004ab} and a Zeeman spin constraint method were used
to constraint the non-collinear spin configurations.
Although the penalty-function spin constraint can cause a small error of spin directions, the error can be less than 0.1 {\%} from the intended spin direction.
We used RMM-DIISH \citep{kresse1996efficient} mixing scheme within the local spin density functional of Ceperley-Alder (LSDA-CA) \citep{ceperley1980ground}. The RMM-DIISH is a suitable mixing scheme for non-collinear spin calculations. In non-collinear spin calculation,
LSDA-CA tends to require fewer K-points than
generalized gradient approximation (GGA) to reduce the error in the determination of spin directions.
We used a combination of RMM-DIISH mixing and LSDA-CA functional for better efficiency.
To describe the on-site Coulomb interaction for Cr atoms, we used the LDA+$U$ method \citep{liechtenstein1995density}.
The effective $U$=3.7eV is adopted for Cr $d$ orbitals.
The value is taken from previous LiCrO$_2$ study \citep{toth2016electromagnon} and the material project data \citep{jain2013commentary}.

\begin{table}[b]
\caption{\label{tab:FM_AFM}%
Total energies for different magnetic configurations.
The 120$^\circ$ AFM with staggerd chirality structure is determined to be the ground state, while the configuration with a straight chirality has 0.22 meV larger energy.
Other collinear models have much higher energies regardless their direction or multi-layer ordering structures.
}
\begin{ruledtabular}
\begin{tabular}{ll}
 & meV/Cr atom
\\
\hline
AFM staggered chirality &  0.00 \\
AFM $xy$ easy plane, staggered chirality  & 0.014 \\
AFM straight chirality  &  0.22\\
AFM $xy$ easy plane, straight chirality  & 0.23 \\
AFM collinear &  27.88   \\
FM collinear  &  27.06\\
\end{tabular}
\end{ruledtabular}
\end{table}

\section{Results}

\subsection{Optimized crystal structure}

First, we optimize the unit cell with the OpenMX package.
Starting from the experimental cell parameters $a=2.923${\AA}, $c=18.08$ {\AA} and internal coordinate $z=0.1105$ described in \citep{2018_DucLe_PdCrO2}, we obtained the DFT-optimized cell parameters: $a=2.889$ {\AA}, $c=17.867$ {\AA} and $z=0.1099$.
The slightly underestimated cell volume is known to be a typical case expected from the use of the LDA exchange-correlation functional. In contrast, the agreement of the internal coordinate $z$ is remarkable.
Fig. \ref{fig:structure} illustrates the triclinic unit cell, which we choose for our DFT calculations.
The conventional hexagonal unit cell requires six layers to describe even number periodicity of staggered chirality.
The 6-layer supercell has an advantage for describing various kinds of 6-layer structures,
but the inefficiency of a larger cell size makes the high precision calculations difficult practically.
Therefore, we adopted a minimal cell, which can describe the directional degree of freedom for all Cr local magnetic moments with the layer and the even number periodicity along the $z$-direction.

Given the optimized unit cell, we calculate total energies for all the relevant spin configurations and analyze them by introducing effective spin models.
We also present Fermi surfaces, which is dependent on the magnetic ordering.
To describe the $k_z=0$ plane, we use $\bm{b}_1$ and $\bm{b}_2+\frac { 2 }{ 3 } \bm{b}_3$ vectors, 
instead the reciprocal vectors $\bm{b}_1$ and $\bm{b}_2$, because the reciprocal lattice vectors of the triclinic unit cell do not have $C_3$ symmetry.

\begin{figure}
\includegraphics[width=0.5\textwidth]{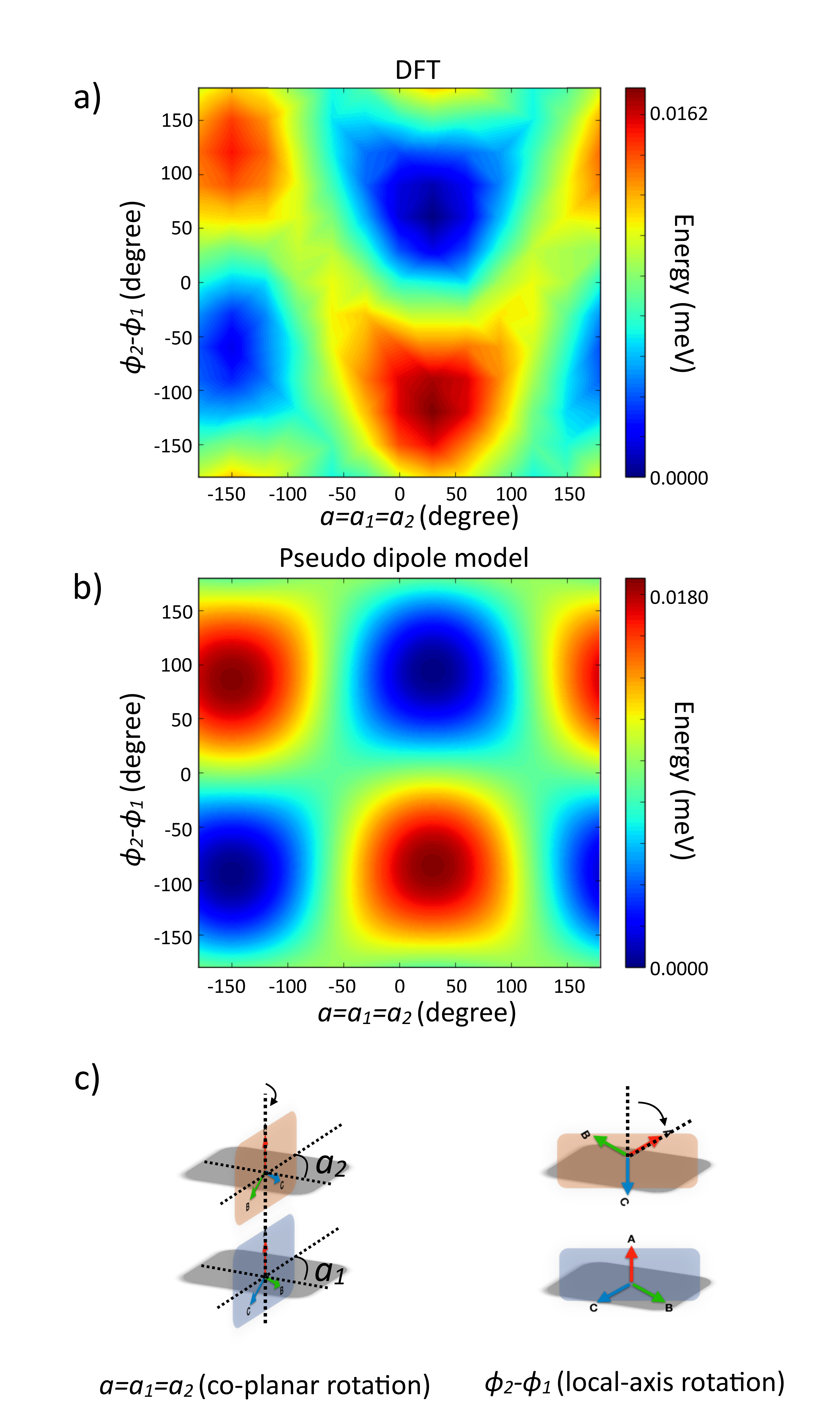}
\caption{\label{fig:spinmodel1}
Easy-plane ($\alpha=\alpha_1=\alpha_2$) and local-easy-axis ($\phi=\phi_2-\phi_1$) rotations calculated in DFT and pseudo dipole model.
a) The ground state of 2-layers model in DFT calculation is  $\alpha=\alpha_1=\alpha_2=30^\circ$ and $\phi=\phi_2-\phi_1=60^\circ$.
b) Pseudo dipole model calculation.
 The ground state in pseudo dipole interaction is $\alpha=30^\circ$ and $\phi=94^\circ$.
c) Schematic figures of co-planar easy-plane rotation (left) and local-easy-axis rotation (right).
}
\end{figure}

\subsection{Magnetic structures}

As shown in Table~\ref{tab:FM_AFM}, the ground-state magnetic configuration has a 120$^\circ$ AFM ordering in the layer, staggered chirality structure along $z$-direction and high symmetric co-planar easy-planes and local-easy-axis (Fig.~\ref{fig:closing}).
The ground magnetic structure can be described by the expression suggested in Ref. \citep{2014_Takatsu_PdCrO2}:

\begin{eqnarray}
\label{eq:equation 1}
{ \bm{S} }_{ { A }_{ n } }&=&S\left[ \bm{ \hat { z } } \cos { { \phi  }_{ n } } +{\bm{ \hat { e } } }_{ \alpha  }\sin { { \phi  }_{ n } }  \right],
\nonumber\\ { \bm{S} }_{ { B }_{ n } }&=&S\left[ \bm{\hat { z }} \cos { \left( { \phi  }_{ n }+\xi _{ n }\frac { 2\pi  }{ 3 }  \right)  } +{\bm{ \hat { e } } }_{ \alpha  }\sin { \left( { \phi  }_{ n }+\xi _{ n }\frac { 2\pi  }{ 3 }  \right)  }  \right] ,
\nonumber\\ { \bm{S} }_{ { C }_{ n } }&=&S\left[ \bm{\hat { z }} \cos { \left( { \phi  }_{ n }-\xi _{ n }\frac { 2\pi  }{ 3 }  \right)  } +{ \bm{\hat { e }}  }_{ \alpha  }\sin { \left( { \phi  }_{ n }-\xi _{ n }\frac { 2\pi  }{ 3 }  \right)  }  \right] ,
\nonumber\\ { \bm{\hat { e }}  }_{ \alpha  }&=&\bm{\hat { x }} \cos { \alpha  } +\bm{\hat { y }} \sin { \alpha  }.
\nonumber\\
\end{eqnarray}
$\bm{S}_{A_n}$, $\bm{S}_{B_n}$ and $\bm{S}_{C_n}$ are local magnetic moments of atomic site $A_n$, $B_n$ and $C_n$ where $n$ is the layer index.
The easy-plane contains $\bm{\hat{z}}$ and $\bm{\hat{e}_{\alpha}}$ vectors, so the azimuthal angle $\alpha$ determines the easy-plane direction.
We can change $\bm{\hat{z}}$ and $\bm{\hat{e}_{\alpha}}$ to another orthogonal set of two unit-vectors so that we can define the easy-plane which does not contain the $\bm{\hat{z}}$ vector (e.g. $xy$-plane).
$\phi_n$ is a local-axis rotation angle inside the easy-plane. $\xi_n$ is $\pm1$, which represents the chirality of spin rotation.
However, the inversion of chirality can also be described by $\alpha \rightarrow  \alpha +180^\circ$.
That means that the straight and staggered chirality is special cases of twist of two easy-planes ($\alpha_2 = \alpha_1 +180^\circ$).

Table~\ref{tab:FM_AFM} lists the total energies of the collinear FM and AFM models, $xy$-easy-plane models, and straight chirality models.
The collinear AFM cannot be stabilized in a layer, so we use 2-layer collinear AFM, which consists of two FM layers pointing the opposite directions.
In this table, we find that the magnetic structure has three different energy contributions.
The collinear spin ordering of either FM or AFM gives rise to the high energy of approximately 27 meV / Cr atom, which is consistent with the Heisenberg exchange interaction.
However, the most intriguing points are its dependence on the chirality (twisting mode of easy-planes) and co-planar easy-planes rotations.
The energy contribution from chirality is approximately 0.22 meV / Cr atom, and that of co-planar easy-planes rotating is 0.017meV / Cr atom.
While the energy scale of these rotating modes is so small, compared to the robust exchange terms, the rotating modes rarely affect the AFM-FM tilting energy, and vice versa.
The more interesting point is that the co-planar rotation mode of easy-planes does not affect to chirality (or twisting of the easy-planes) either.
It indicates that twisting mode and, co-planar easy-planes and local-axis rotating mode have separated origins.

We may consider three separated contributions with different energy scales to the magnetic interactions, which can be described by three separate effective spin models.

\begin{table}[b]
\caption{\label{tab:PDOS}%
The proportion of orbitals at Fermi level. calculated in 20 $\times$ 20 $\times$ 20 k-points, with spin parameters in Figure. \ref{fig:DFT}.
As previous PdCrO$_2$ studies have expected, Half-filled Pd $d_{z^2}$ is major Fermi surface constructing orbital (See Figure.\ref{fig:DFT}.c)), and other pd $d$ components are bonding orbitals
However there are still large proportion of $d_{x^{2}-y^{2}}$ and $d_{xy}$ orbitals at Fermi level that might affect tight binding hopping models in Figure.\ref{fig:TB}. $1.8\%$ of Cr $d_{z^2}$ orbital is small amount, but that could be a key feature of band structures of DFT and tight binding models, also $z$-directional spin interactions. 
}
\begin{ruledtabular}
\begin{tabular}{lrlrlr}
Pd            & 82.61\%   & Cr            & 4.85\%  & O       & 12.54\% \\
Pd $d_{z^2}$    & 33.13\%   & Cr $d_{z^2}$    & 1.81\%  & O $p_{x}$ & 1.36\% \\
Pd $d_{x^2y^2}$ & 18.22\%   & Cr $d_{x^2y^2}$ & 0.33\%  & O $p_{y}$ & 1.32\% \\
Pd $d_{xy}$     & 18.22\%   & Cr $d_{xy}$     & 0.31\%  & O $p_{z}$ & 9.16\% \\
Pd $d_{xz}$     & 0.44\%    & Cr $d_{xz}$     & 0.56\%  &         &     \\
Pd $d_{yz}$     & 0.44\%    & Cr $d_{yz}$     & 0.60\%  &         &     \\
\end{tabular}
\end{ruledtabular}
\end{table}

\begin{figure}
\includegraphics[width=0.40\textwidth]{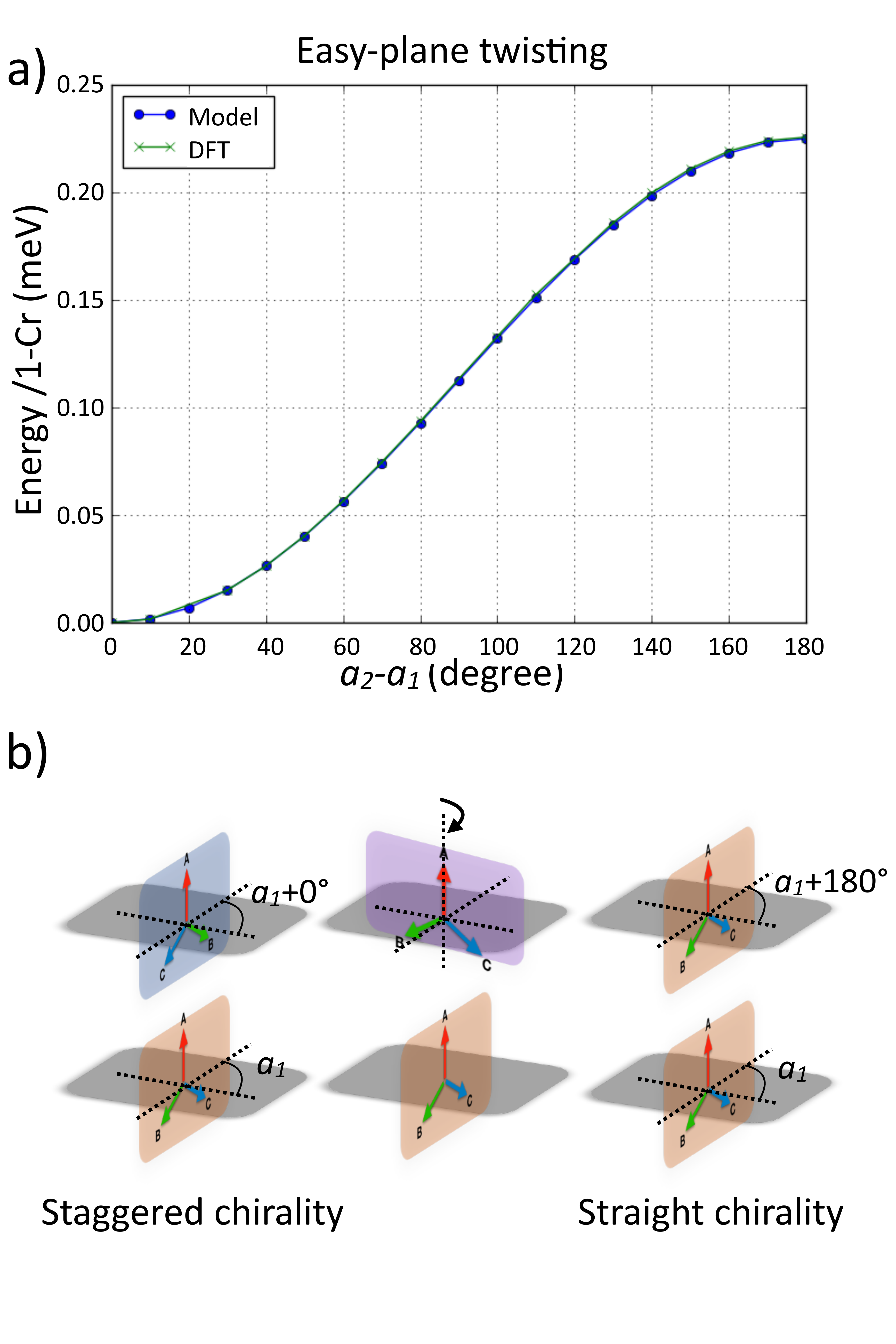}
\caption{\label{fig:spinmodel2}
Twisting of easy-planes and its energy.
The twisting easy-planes between upper layer($\alpha_2$) and lower layer($\alpha_1$) makes a chirality reverse when $180^\circ$ $\alpha_2 - \alpha_1 = 180^\circ$.
a)
Green line :  DFT energy.
blue line : Cyclic 4-spins interaction model with the parameter $J_{ring}=0.1$meV.
DFT and the model have nearly same curves.
b) Schematic figures of twisted easy-planes.
}
\end{figure}

\subsection{\label{sec:spinmodel}Effective spin models \protect\\}

Let us consider Heisenberg terms which can account for the energy difference between FM and AFM configurations as obtained from DFT calculations.

\begin{align}
H_1={ J }_{ 1 }\sum _{ \left< i,j \right>  }^{  }{ { \bm{S} }_{ i } } \cdot { \bm{S} }_{ j }+{ J }_{ 2 }\sum _{ \left< \left< i,j \right>  \right>  }^{  }{ { \bm{S} }_{ i } } \cdot { \bm{S} }_{ j }\nonumber\\+{ J }_{ 3 }\sum _{ \left< \left< \left< i,j \right>  \right>  \right>  }^{  }{ { \bm{S} }_{ i } } \cdot { \bm{S} }_{ j }+\quad \cdots 
\end{align}{\label{eq:Heisenberg}}

However, the Heisenberg terms of Eq.~(2) do not contain any contribution to chirality, easy-plane, and local-easy-axis directions.
Also, Dzyaloshinskii-Moriya (DM) interaction is forbidden by symmetry \citep{2018_DucLe_PdCrO2,moriya1960anisotropic}.
First, we show that the 4-cyclic ring interaction of nearest neighbor spins can be a good effective model for chirality and twisting easy-plane mode.

It has been challenging to make an effective model for describing easy-plane twisting or chirality.
Because previous 2-spins interaction models do not separate those twisting modes from FM tilting or co-planar easy-plane rotation and local-axis rotation, the energy contribution from straight and staggered chirality must contain a product term of two spin rotation parameter $\xi_n$. Also, it has to be separated from $\phi_n$ and co-planar ($\alpha_n=\alpha_{n+1}$) rotation mode.
To satisfy the two conditions, we suppose a product of relative spin directions of each layer.
To describe the relative spin direction, we need at least two spins in a layer.
Therefore, the minimum description of our suggestion should be the interactions between two-spin pairs, which has a total of four spins.
Subsequently, we assume the nearest interactions, which is the simplest form within the first assumption.
Then, the selected four nearest-neighbor spins can form a cyclic loop.
Finally, we assume the interaction is connected to the hopping of electrons among local Cr spins so that we can represent each local magnetic moment by spinor.

Our suggested effective model can be represented as follows.
\begin{align}
H_{cyclic}&={ J }_{ ring }\sum _{ i }^{\substack{cyclic-\\rings}}\left< { S }_{ i }|{ S }_{ j } \right> \left< { S }_{ j }|{ S }_{ k  } \right> \left< { S }_{ k  }|{ S }_{ l  } \right> \left< { S }_{ l  }|{ S }_{ i } \right>
\nonumber\\
\left| { S }_{ n } \right> &=\left( \begin{matrix} \cos { \frac { { \theta  }_{ n } }{ 2 }  }  \\ \sin { \frac { { \theta  }_{ n } }{ 2 }  } { e }^{ i{ \varphi  }_{ n } } \end{matrix} \right) 
\end{align}
where $S_i,S_j,S_k,S_l$ is a sequence of Cr local spins in the 1st nearest cyclic loop.
Not only it has an energy scale far from Heisenberg interaction energy, it is also separated from co-planar easy-planes rotating and local-axis rotating, which are the third factor in DFT calculations.
We can interpret that each local spin is a hopping site, and they also construct projection matrices for spins.

Fig.~\ref{fig:spinmodel2} shows continuous change of easy-plane twisting ($\alpha = \alpha_2-\alpha_1$) and its energetic behavior.
When $\alpha = \alpha_2-\alpha_1 = 180^\circ$, the chirality is reversed.
The calculated energy of the model and DFT perfectly fit each other when $J_{ring}$ = 0.1meV.

Still, there should be another interaction model to describe co-planar easy-plane rotation and local-axis.
Previously, dipole interaction was suggested for a possible candidate of small energy interaction term, but they did not show an energy calculation or neutron scattering fitting for it \citep{2018_DucLe_PdCrO2}.
Dipole interaction is not a good model to rescale its energy, so we set pseudo dipole interaction term as follows

\begin{align}
\label{eq:dipole}
{ H }_{ ij }^{ dip }={ \bm{S} }_{ i }{ \bar{\bar { \bm{A} }}  }_{ ij }{ \bm{S} }_{ j }\quad ,
\nonumber\\ { \bar{\bar{ \bm{A} } } }_{ ij }={ D }_{ ij }\left( 3{ \bm{\hat { r }}  }_{ ij }{ \bm{\hat { r }}  }_{ ij }^{ \top  }-{ \delta  }_{ ij } \right) 
\end{align}
where ${ \bm{\hat { r }}  }_{ ij }$ is local easy direction between atoms at $i_{th}$ and $j_{th}$ sites.
In Fig. \ref{fig:spinmodel1}, both DFT and pseudo dipole model have $\alpha_1 = \alpha_2 = 30^\circ$ as a ground state.
It is the direction contains 1st nearest inter layer connections.
However DFT calculation shows minimum energy at
$\phi=\phi_2-\phi_1=60^\circ$, while pseudo dipole model has minimum energy at $\phi=\phi_2-\phi_1=94^\circ$.

\begin{figure}
\includegraphics[width=0.4\textwidth]{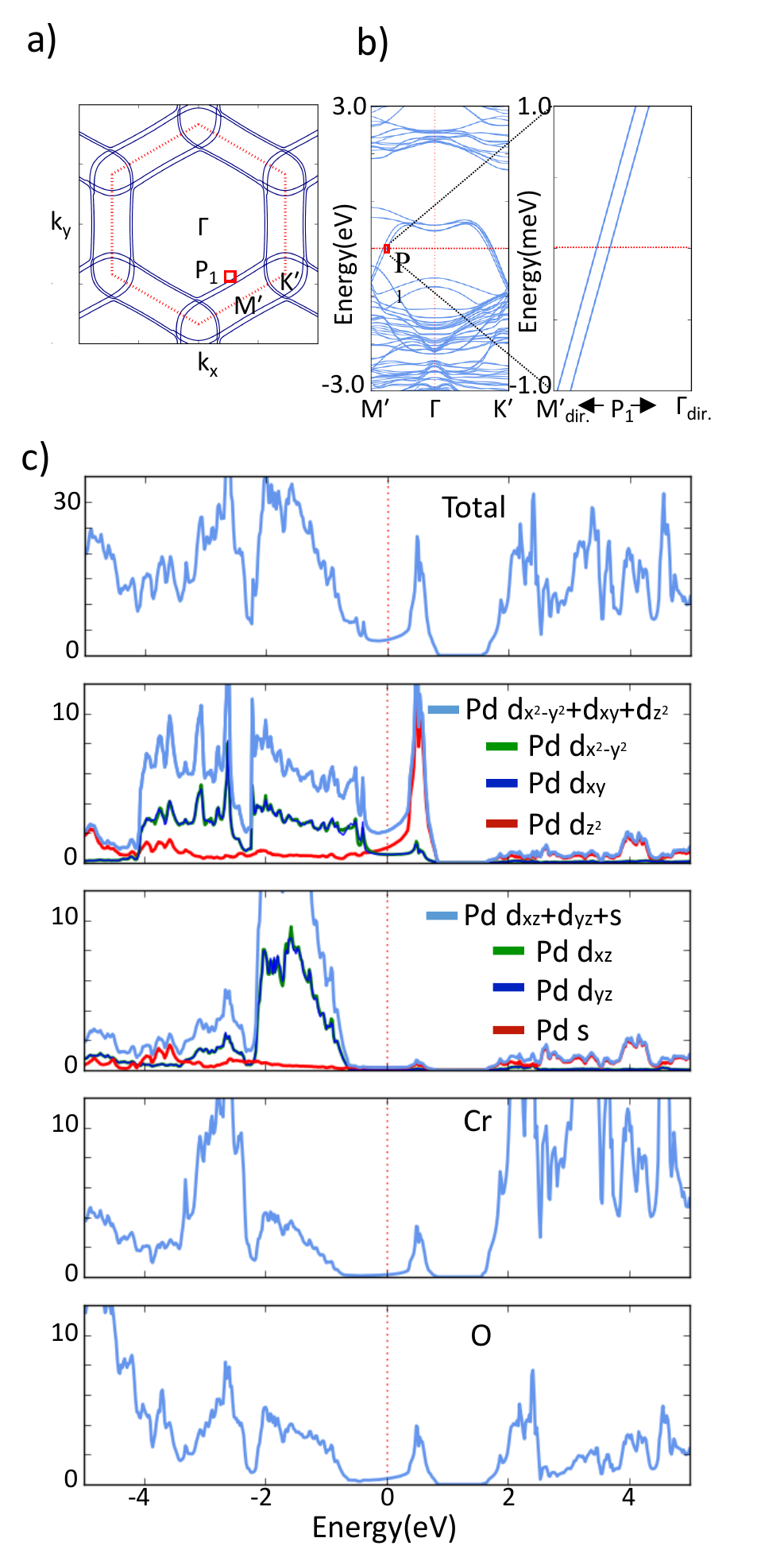}
\caption{\label{fig:DFT} 
PdCrO$_2$ electronic bands structure and PDOS.
a) A Fermi surface at $k_z=0$ plane.
Since the unit cell, which we used to describe staggered chirality in DFT calculation, has inclined a$_3$ lattice.
We used $\bm{b}_2 + \frac { 2 }{ 3 } \bm{b}_3$ instead $\bm{b}_2$ to illustrate $k_z=0$ plane.
Points in Figure defined as $M'=\frac{1}{2}\bm{b}_1$ and $K'=\frac{2}{3}\bm{b}_1 +\frac{1}{3}\bm{b}_2 +\frac{2}{9}\bm{b}_3$.
$P_1$ is the 1st touching band of the line from $\Gamma$ to $M'$.
b) Band structure in $M'-\Gamma-K'$ line and $P_1$ zoom-in view.
c) DOS and PDOS.
See Table \ref{tab:PDOS} for details of PDOS at the Fermi level.
}
\end{figure}

\subsection{Fermi surfaces}

\begin{figure}
\includegraphics[width=0.4\textwidth]{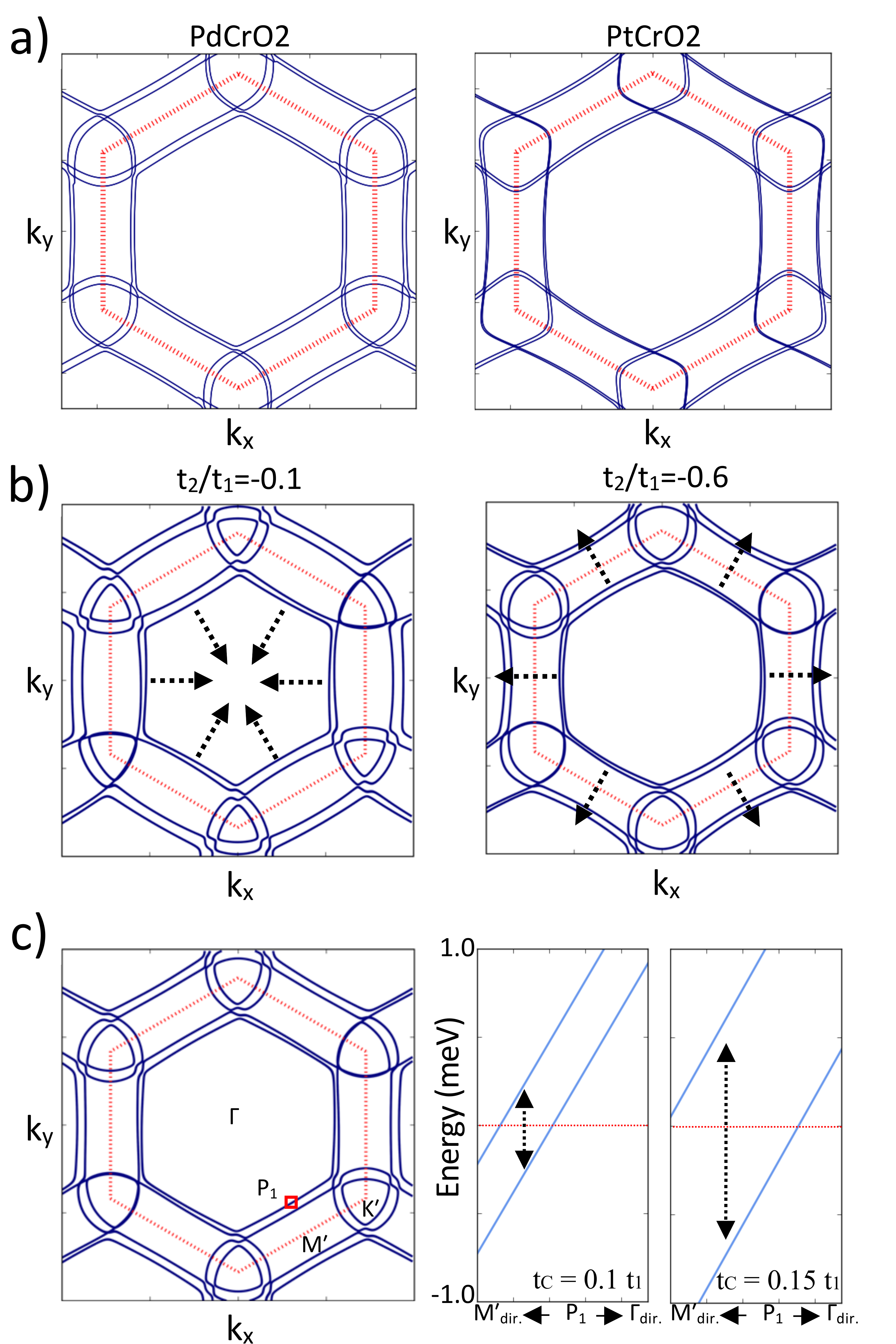}
\caption{\label{fig:TB} 
DFT and TB calculations of Fermi surfaces at $k_z$=0 plane.
a) DFT calculation of PdCrO$_2$ (left) and PtCrO$_2$ (right)
b),c) TB calculations.
b) Controlled Fermi surface shapes by second nearest direct hopping parameter $t_2$.
$t_2 = -0.1t_1$ (left), $t_2 = -0.6t_1$ (right).
c) Controlled degeneracy breaking by hopping parameter $t_c$.
$t_c=0.1t_1$ (middle) and $t_c=0.15t_1$ with $t_2=-0.35t_1$.
}
\end{figure}

Fig \ref{fig:DFT} shows an electronic structure calculated in DFT with an example parameters $\alpha_1 =31, \alpha_2=31, \phi_1=17, \phi_2=16, \xi_1=1, \xi_2=-1$.
The Fermi surface is nearly hexagonal shape on $k_z=0$ plane, and it has very weak $z$-directional dispersion.
Pd $d$ electrons are the major component at the Fermi level.
Pd $s$ accounts for less than $3\%$ of sum of Pd $d$ electrons.
Cr $d_{z^2}$ accounts only for $1.8\%$.
It also implies that $z$-directional Pd-Cr hopping is weak.
One intriguing result is that there is small degeneracy breaking on the bands near the Fermi level.
The degeneracy breaking is approximately 0.1 meV.
To study $A$-site $d$ orbital effects, we also calculated PtCrO$_2$.
Fig \ref{fig:TB} a) shows DFT calculation of Fermi surface of PtCrO$_2$.
The Fermi surface of PtCrO$_2$ is convex (concave in 2nd zone).
This result is similar to the Fermi surface of PtCoO$_2$, which is observed in ARPES \citep{kushwaha2015nearly}.
The portion of Cr $d$ electron is very small at Fermi level, but it is still essential to describe the Fermi surface.

\begin{figure}
\includegraphics[width=0.35\textwidth]{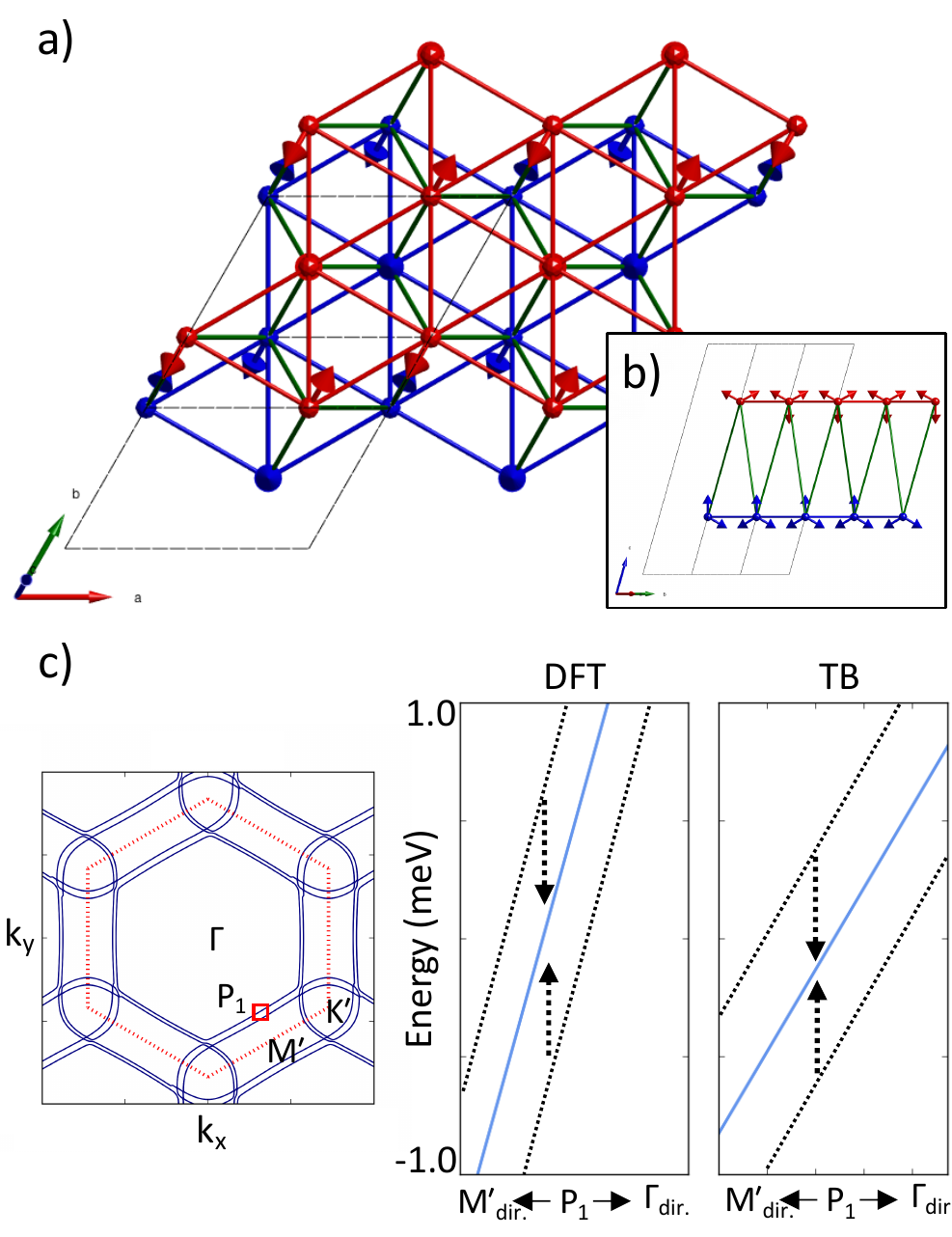}
\caption{\label{fig:closing}
a),b) The ground magnetic structure calculated in DFT.
c) The Fermi surface of DFT calculation (left), degenerated 2 bands calculated in DFT (middle) and TB model (right).
}
\end{figure}

To figure out the factor which changes Fermi surface shape and the tiny structures in electronic bands, we calculate TB models.
For simplicity, we assumed that only Pd atoms have hopping sites (half filled) and Cr atoms have local magnetic moments which construct projection matrices for Pd electrons.
First, the TB model is simplified to be a quasi two dimensional triangular lattice model.
By assuming Pd $d$ electrons have 1st nearest and 2nd nearest hoppings along $xy$-plane directions, we constructed an effective Hamiltonian

\begin{eqnarray}
H_{0,l}=t_1\sum _{ \left< i,j \right> l \sigma  }^{  }{ { d }_{ il\sigma  }^{ \dagger  }{ d }_{ jl\sigma  } } +t_2\sum _{ \left< \left< i,j \right>  \right> l\sigma  }^{  }{ { d }_{ il\sigma  }^{ \dagger  }{ d }_{ jl\sigma  } } 
\end{eqnarray}
where $i,j$ are hopping sites indices, $\sigma$ is a spin index and $l$ is a layer index.
The nearest neighbor hopping $t_1$ and 2nd nearest neighbor hopping $t_2$ determine $xy$-planar shape of the Fermi surface.
At $t_2=-0.35t_1$, the Fermi surface is getting closer to a flat hexagonal shape.
However, when $t_2$ changes, the shape of Fermi surface are getting closer to concave or convex shapes.
When $t_2=-0.1t_1$, a hexagonal lines become concave (convex on 2nd zone), and when $t_2=-0.6t_1$, a hexagonal lines become convex (concave on 2nd zone).
In this model, the 1st and 2nd hopping ratio of $A$-site of $AB$O$_2$ delaffosite is the parameter which controls Fermi surface shape.

Meanwhile, Cr atoms contribute to only small energy, but they construct $z$-directional connections and tiny structures on bands. 
We set up Pd-Cr-Pd interaction which includes inter layer hoppings.
\begin{align}
H_{1,ll'}&=t_c\sum _{ i,j',l,l',\sigma ,\sigma ',n }^{  }{ \left< \sigma |{ S }_{ n } \right> \left< { S }_{ n }|{ \sigma  }' \right> { d }_{ il\sigma  }^{ \dagger  }{ d }_{ j'l'\sigma'  }  } 
\nonumber\\ \left| { S }_{ n } \right> &=\quad \left( \begin{matrix} \cos { \frac { { \theta  }_{ n } }{ 2 }  }  \\ \sin { \frac { { \theta  }_{ n } }{ 2 }  } { e }^{ i{ \varphi  }_{ n } } \end{matrix} \right) 
\end{align}
$S_n$ is a non-collinear local Cr spin moment in Pd-Cr-Pd hopping path (nearest Pd-Cr hopping path).
The moments on hopping path are interpreted as a projection matrix $ \left| {S }_{ n } \right> \left<{ S }_{ n }\right|$.
It creates off-diagonal term for Hamiltonian matrix, breaks degeneracy of bands.
Because magnetic projection hopping terms are determined by magnetic structures, the tiny electronic structure is controlled by magnetic structures, specifically easy-plane and local-axis directions.
In Fig.~\ref{fig:closing}, it show the model describes well the degeneracy breaking in DFT.
At the high symmetric magnetic structure ($\alpha=30^\circ$,$\phi_1=0^\circ$ and $\phi_2=60^\circ$), which has the minimum energy in DFT calculation, both DFT and TB calculation show restoring of degeneracy. Therefore, although the portion of Cr is small on Fermi level, and the energy from easy-plane and local-axis rotation mode is very small, that tiny structure is very important to describe how $z$-directional interaction can be constructed and how the band degeneracy breaking can be controlled.

\section{\label{sec:level1}Conclusion \protect\\ }

In this work, we studied the electronic and magnetic structures of PdCrO$_2$ with the first-principles calculations and model calculations.
We used spin constraint DFT calculation method to understand the energetic behavior and the Fermi surface changes with related magnetic structures.

The ground magnetic structure is AFM staggered chirality with high symmetric easy-plane and easy-axis ($\xi_1=+1,\xi_2=-1, \alpha = 30^\circ, \phi_1 = 0^\circ \phi_2 = 60^\circ$).
There are 3 different factors which dominate each energetic region of magnetic structures.
Consequently, we suggest three effective spin models which contribute to each energy range  separately.
The strongest magnetic ordering energy comes from AFM and FM ordering, and Heisenberg interaction model easily predict their behavior.
Chirality (or twisting ease-planes mode) is the second factor, approximately 1/100 of the Heisenberg exchange interaction.
We suggest that the cyclic 4-spin ring interaction model effectively fit DFT result of twisting easy-planes, also it is independent from the first and third factor.
The energy from co-planar ease-plane rotating and local-axis rotating mode is the 3rd factor, they gives very small energy, approximately 1/1000 of the Heisenberg interaction.
Pseudo dipole interaction model predict the same easy-plane direction with the DFT calculation, but it has ground state at $\phi=94^\circ$.
Still, the rotating mode of easy-plane and local-axis is a difficult problem in both experimental data fittings and DFT calculations.
Furthermore, there could be much more complicated details on easy-plane and local-axis directions, because they can also have 6-layer, 12-layer and more multi-layer periodicity.
Therefore, there should be multi-layer supercell calculations to examine further details on them.

In electronic structures, we found magnetic structure dependent Fermi surfaces.
The nearly hexagonal shape of Fermi surface have a weak $z$-directional connection.
It also have tiny degeneracy breaking, which is approximately 0.1meV.
Pd $d$ electron is the major component of it, while Cr $d$ electrons take only small portion.
However, magnetic moments of Cr atoms are very important to explain electronic structure of this system.
By setting up the first nearest and second nearest hopping TB model, we found that shape of Fermi surface is primarily controlled by $A$-site atoms (Pd or Pt). 
The result is consistent with experimental data and DFT calculation, it explains why Pt-delafossite materials have concave hexagonal shape of Fermi surfaces.
Meanwhile, Cr local magnetic moments can be a bridge between Pd inter layers.
Besides, they can be interpreted as perturbative spin projection terms.
The Pd-Cr-Pd magnetic hopping model describes how the tiny degeneracy breaking can be constructed in the DFT calculation.
Magnetic-structure-controlled degeneracy breaking might be correlated to AHE, since it changes very small energy gap between occupied band and unoccupied band \citep{wang2006ab}.
Therefore, specific easy-plane, local-axis structures and their responses to external field can be the key to understand AHE in PdCrO$_2$ and related delafossite magnetic systems.

\section*{Acknowledgements}

We gratefully acknowledge Prof. Je-Geun Park and Dr. M. Duc Le for valuable help and discussions.
This work was supported by the National Research Foundation of Korea (NRF) grant funded by the Korea government (no. 2017R1A2B4007100).


\begin{thebibliography}{32}%
\makeatletter
\providecommand \@ifxundefined [1]{%
 \@ifx{#1\undefined}
}%
\providecommand \@ifnum [1]{%
 \ifnum #1\expandafter \@firstoftwo
 \else \expandafter \@secondoftwo
 \fi
}%
\providecommand \@ifx [1]{%
 \ifx #1\expandafter \@firstoftwo
 \else \expandafter \@secondoftwo
 \fi
}%
\providecommand \natexlab [1]{#1}%
\providecommand \enquote  [1]{``#1''}%
\providecommand \bibnamefont  [1]{#1}%
\providecommand \bibfnamefont [1]{#1}%
\providecommand \citenamefont [1]{#1}%
\providecommand \href@noop [0]{\@secondoftwo}%
\providecommand \href [0]{\begingroup \@sanitize@url \@href}%
\providecommand \@href[1]{\@@startlink{#1}\@@href}%
\providecommand \@@href[1]{\endgroup#1\@@endlink}%
\providecommand \@sanitize@url [0]{\catcode `\\12\catcode `\$12\catcode
  `\&12\catcode `\#12\catcode `\^12\catcode `\_12\catcode `\%12\relax}%
\providecommand \@@startlink[1]{}%
\providecommand \@@endlink[0]{}%
\providecommand \url  [0]{\begingroup\@sanitize@url \@url }%
\providecommand \@url [1]{\endgroup\@href {#1}{\urlprefix }}%
\providecommand \urlprefix  [0]{URL }%
\providecommand \Eprint [0]{\href }%
\providecommand \doibase [0]{http://dx.doi.org/}%
\providecommand \selectlanguage [0]{\@gobble}%
\providecommand \bibinfo  [0]{\@secondoftwo}%
\providecommand \bibfield  [0]{\@secondoftwo}%
\providecommand \translation [1]{[#1]}%
\providecommand \BibitemOpen [0]{}%
\providecommand \bibitemStop [0]{}%
\providecommand \bibitemNoStop [0]{.\EOS\space}%
\providecommand \EOS [0]{\spacefactor3000\relax}%
\providecommand \BibitemShut  [1]{\csname bibitem#1\endcsname}%
\let\auto@bib@innerbib\@empty
\bibitem [{\citenamefont {Takatsu}\ \emph {et~al.}(2009)\citenamefont
  {Takatsu}, \citenamefont {Yoshizawa}, \citenamefont {Yonezawa},\ and\
  \citenamefont {Maeno}}]{2009_Takatsu_PdCrO2}%
  \BibitemOpen
  \bibfield  {author} {\bibinfo {author} {\bibfnamefont {H.}~\bibnamefont
  {Takatsu}}, \bibinfo {author} {\bibfnamefont {H.}~\bibnamefont {Yoshizawa}},
  \bibinfo {author} {\bibfnamefont {S.}~\bibnamefont {Yonezawa}}, \ and\
  \bibinfo {author} {\bibfnamefont {Y.}~\bibnamefont {Maeno}},\ }\href@noop {}
  {\bibfield  {journal} {\bibinfo  {journal} {Physical Review B}\ }\textbf
  {\bibinfo {volume} {79}},\ \bibinfo {pages} {104424} (\bibinfo {year}
  {2009})}\BibitemShut {NoStop}%
\bibitem [{\citenamefont {Takatsu}\ \emph
  {et~al.}(2010{\natexlab{a}})\citenamefont {Takatsu}, \citenamefont
  {Yonezawa}, \citenamefont {Fujimoto},\ and\ \citenamefont
  {Maeno}}]{2010_Takatsu_PdCrO2}%
  \BibitemOpen
  \bibfield  {author} {\bibinfo {author} {\bibfnamefont {H.}~\bibnamefont
  {Takatsu}}, \bibinfo {author} {\bibfnamefont {S.}~\bibnamefont {Yonezawa}},
  \bibinfo {author} {\bibfnamefont {S.}~\bibnamefont {Fujimoto}}, \ and\
  \bibinfo {author} {\bibfnamefont {Y.}~\bibnamefont {Maeno}},\ }\href@noop {}
  {\bibfield  {journal} {\bibinfo  {journal} {Physical review letters}\
  }\textbf {\bibinfo {volume} {105}},\ \bibinfo {pages} {137201} (\bibinfo
  {year} {2010}{\natexlab{a}})}\BibitemShut {NoStop}%
\bibitem [{\citenamefont {Ok}\ \emph {et~al.}(2013)\citenamefont {Ok},
  \citenamefont {Jo}, \citenamefont {Kim}, \citenamefont {Shishidou},
  \citenamefont {Choi}, \citenamefont {Noh}, \citenamefont {Oguchi},
  \citenamefont {Min},\ and\ \citenamefont {Kim}}]{2013_JongMokOk_PdCrO2}%
  \BibitemOpen
  \bibfield  {author} {\bibinfo {author} {\bibfnamefont {J.~M.}\ \bibnamefont
  {Ok}}, \bibinfo {author} {\bibfnamefont {Y.}~\bibnamefont {Jo}}, \bibinfo
  {author} {\bibfnamefont {K.}~\bibnamefont {Kim}}, \bibinfo {author}
  {\bibfnamefont {T.}~\bibnamefont {Shishidou}}, \bibinfo {author}
  {\bibfnamefont {E.}~\bibnamefont {Choi}}, \bibinfo {author} {\bibfnamefont
  {H.-J.}\ \bibnamefont {Noh}}, \bibinfo {author} {\bibfnamefont
  {T.}~\bibnamefont {Oguchi}}, \bibinfo {author} {\bibfnamefont
  {B.}~\bibnamefont {Min}}, \ and\ \bibinfo {author} {\bibfnamefont {J.~S.}\
  \bibnamefont {Kim}},\ }\href@noop {} {\bibfield  {journal} {\bibinfo
  {journal} {Physical review letters}\ }\textbf {\bibinfo {volume} {111}},\
  \bibinfo {pages} {176405} (\bibinfo {year} {2013})}\BibitemShut {NoStop}%
\bibitem [{\citenamefont {Takatsu}\ \emph {et~al.}(2014)\citenamefont
  {Takatsu}, \citenamefont {N{\'e}nert}, \citenamefont {Kadowaki},
  \citenamefont {Yoshizawa}, \citenamefont {Enderle}, \citenamefont {Yonezawa},
  \citenamefont {Maeno}, \citenamefont {Kim}, \citenamefont {Tsuji},
  \citenamefont {Takata} \emph {et~al.}}]{2014_Takatsu_PdCrO2}%
  \BibitemOpen
  \bibfield  {author} {\bibinfo {author} {\bibfnamefont {H.}~\bibnamefont
  {Takatsu}}, \bibinfo {author} {\bibfnamefont {G.}~\bibnamefont {N{\'e}nert}},
  \bibinfo {author} {\bibfnamefont {H.}~\bibnamefont {Kadowaki}}, \bibinfo
  {author} {\bibfnamefont {H.}~\bibnamefont {Yoshizawa}}, \bibinfo {author}
  {\bibfnamefont {M.}~\bibnamefont {Enderle}}, \bibinfo {author} {\bibfnamefont
  {S.}~\bibnamefont {Yonezawa}}, \bibinfo {author} {\bibfnamefont
  {Y.}~\bibnamefont {Maeno}}, \bibinfo {author} {\bibfnamefont
  {J.}~\bibnamefont {Kim}}, \bibinfo {author} {\bibfnamefont {N.}~\bibnamefont
  {Tsuji}}, \bibinfo {author} {\bibfnamefont {M.}~\bibnamefont {Takata}},
  \emph {et~al.},\ }\href@noop {} {\bibfield  {journal} {\bibinfo  {journal}
  {Physical Review B}\ }\textbf {\bibinfo {volume} {89}},\ \bibinfo {pages}
  {104408} (\bibinfo {year} {2014})}\BibitemShut {NoStop}%
\bibitem [{\citenamefont {Noh}\ \emph {et~al.}(2014)\citenamefont {Noh},
  \citenamefont {Jeong}, \citenamefont {Chang}, \citenamefont {Jeong},
  \citenamefont {Moon}, \citenamefont {Cho}, \citenamefont {Ok}, \citenamefont
  {Kim}, \citenamefont {Kim}, \citenamefont {Min} \emph
  {et~al.}}]{2014_HanJinNoh_PdCrO2}%
  \BibitemOpen
  \bibfield  {author} {\bibinfo {author} {\bibfnamefont {H.-J.}\ \bibnamefont
  {Noh}}, \bibinfo {author} {\bibfnamefont {J.}~\bibnamefont {Jeong}}, \bibinfo
  {author} {\bibfnamefont {B.}~\bibnamefont {Chang}}, \bibinfo {author}
  {\bibfnamefont {D.}~\bibnamefont {Jeong}}, \bibinfo {author} {\bibfnamefont
  {H.~S.}\ \bibnamefont {Moon}}, \bibinfo {author} {\bibfnamefont {E.-J.}\
  \bibnamefont {Cho}}, \bibinfo {author} {\bibfnamefont {J.~M.}\ \bibnamefont
  {Ok}}, \bibinfo {author} {\bibfnamefont {J.~S.}\ \bibnamefont {Kim}},
  \bibinfo {author} {\bibfnamefont {K.}~\bibnamefont {Kim}}, \bibinfo {author}
  {\bibfnamefont {B.}~\bibnamefont {Min}},  \emph {et~al.},\ }\href@noop {}
  {\bibfield  {journal} {\bibinfo  {journal} {Scientific reports}\ }\textbf
  {\bibinfo {volume} {4}},\ \bibinfo {pages} {3680} (\bibinfo {year}
  {2014})}\BibitemShut {NoStop}%
\bibitem [{\citenamefont {Billington}\ \emph {et~al.}(2015)\citenamefont
  {Billington}, \citenamefont {Ernsting}, \citenamefont {Millichamp},
  \citenamefont {Lester}, \citenamefont {Dugdale}, \citenamefont {Kersh},
  \citenamefont {Duffy}, \citenamefont {Giblin}, \citenamefont {Taylor},
  \citenamefont {Manuel} \emph {et~al.}}]{2015_DavidBillington_PdCrO2}%
  \BibitemOpen
  \bibfield  {author} {\bibinfo {author} {\bibfnamefont {D.}~\bibnamefont
  {Billington}}, \bibinfo {author} {\bibfnamefont {D.}~\bibnamefont
  {Ernsting}}, \bibinfo {author} {\bibfnamefont {T.~E.}\ \bibnamefont
  {Millichamp}}, \bibinfo {author} {\bibfnamefont {C.}~\bibnamefont {Lester}},
  \bibinfo {author} {\bibfnamefont {S.~B.}\ \bibnamefont {Dugdale}}, \bibinfo
  {author} {\bibfnamefont {D.}~\bibnamefont {Kersh}}, \bibinfo {author}
  {\bibfnamefont {J.~A.}\ \bibnamefont {Duffy}}, \bibinfo {author}
  {\bibfnamefont {S.~R.}\ \bibnamefont {Giblin}}, \bibinfo {author}
  {\bibfnamefont {J.~W.}\ \bibnamefont {Taylor}}, \bibinfo {author}
  {\bibfnamefont {P.}~\bibnamefont {Manuel}},  \emph {et~al.},\ }\href@noop {}
  {\bibfield  {journal} {\bibinfo  {journal} {Scientific reports}\ }\textbf
  {\bibinfo {volume} {5}},\ \bibinfo {pages} {12428} (\bibinfo {year}
  {2015})}\BibitemShut {NoStop}%
\bibitem [{\citenamefont {Le}\ \emph {et~al.}(2018)\citenamefont {Le},
  \citenamefont {Jeon}, \citenamefont {Kolesnikov}, \citenamefont {Voneshen},
  \citenamefont {Gibbs}, \citenamefont {Kim}, \citenamefont {Jeong},
  \citenamefont {Noh}, \citenamefont {Park}, \citenamefont {Yu} \emph
  {et~al.}}]{2018_DucLe_PdCrO2}%
  \BibitemOpen
  \bibfield  {author} {\bibinfo {author} {\bibfnamefont {M.~D.}\ \bibnamefont
  {Le}}, \bibinfo {author} {\bibfnamefont {S.}~\bibnamefont {Jeon}}, \bibinfo
  {author} {\bibfnamefont {A.~I.}\ \bibnamefont {Kolesnikov}}, \bibinfo
  {author} {\bibfnamefont {D.}~\bibnamefont {Voneshen}}, \bibinfo {author}
  {\bibfnamefont {A.}~\bibnamefont {Gibbs}}, \bibinfo {author} {\bibfnamefont
  {J.~S.}\ \bibnamefont {Kim}}, \bibinfo {author} {\bibfnamefont
  {J.}~\bibnamefont {Jeong}}, \bibinfo {author} {\bibfnamefont {H.-J.}\
  \bibnamefont {Noh}}, \bibinfo {author} {\bibfnamefont {C.}~\bibnamefont
  {Park}}, \bibinfo {author} {\bibfnamefont {J.}~\bibnamefont {Yu}},  \emph
  {et~al.},\ }\href@noop {} {\bibfield  {journal} {\bibinfo  {journal}
  {Physical Review B}\ }\textbf {\bibinfo {volume} {98}},\ \bibinfo {pages}
  {024429} (\bibinfo {year} {2018})}\BibitemShut {NoStop}%
\bibitem [{\citenamefont {Lechermann}(2018)}]{lechermann2018hidden}%
  \BibitemOpen
  \bibfield  {author} {\bibinfo {author} {\bibfnamefont {F.}~\bibnamefont
  {Lechermann}},\ }\href@noop {} {\bibfield  {journal} {\bibinfo  {journal}
  {Physical Review Materials}\ }\textbf {\bibinfo {volume} {2}},\ \bibinfo
  {pages} {085004} (\bibinfo {year} {2018})}\BibitemShut {NoStop}%
\bibitem [{\citenamefont {Sobota}\ \emph {et~al.}(2013)\citenamefont {Sobota},
  \citenamefont {Kim}, \citenamefont {Takatsu}, \citenamefont {Hashimoto},
  \citenamefont {Mo}, \citenamefont {Hussain}, \citenamefont {Oguchi},
  \citenamefont {Shishidou}, \citenamefont {Maeno}, \citenamefont {Min} \emph
  {et~al.}}]{2013_Sobota_PdCrO2}%
  \BibitemOpen
  \bibfield  {author} {\bibinfo {author} {\bibfnamefont {J.~A.}\ \bibnamefont
  {Sobota}}, \bibinfo {author} {\bibfnamefont {K.}~\bibnamefont {Kim}},
  \bibinfo {author} {\bibfnamefont {H.}~\bibnamefont {Takatsu}}, \bibinfo
  {author} {\bibfnamefont {M.}~\bibnamefont {Hashimoto}}, \bibinfo {author}
  {\bibfnamefont {S.-K.}\ \bibnamefont {Mo}}, \bibinfo {author} {\bibfnamefont
  {Z.}~\bibnamefont {Hussain}}, \bibinfo {author} {\bibfnamefont
  {T.}~\bibnamefont {Oguchi}}, \bibinfo {author} {\bibfnamefont
  {T.}~\bibnamefont {Shishidou}}, \bibinfo {author} {\bibfnamefont
  {Y.}~\bibnamefont {Maeno}}, \bibinfo {author} {\bibfnamefont {B.~I.}\
  \bibnamefont {Min}},  \emph {et~al.},\ }\href@noop {} {\bibfield  {journal}
  {\bibinfo  {journal} {Physical Review B}\ }\textbf {\bibinfo {volume} {88}},\
  \bibinfo {pages} {125109} (\bibinfo {year} {2013})}\BibitemShut {NoStop}%
\bibitem [{\citenamefont {Ghannadzadeh}\ \emph {et~al.}(2017)\citenamefont
  {Ghannadzadeh}, \citenamefont {Licciardello}, \citenamefont
  {Arsenijevi{\'c}}, \citenamefont {Robinson}, \citenamefont {Takatsu},
  \citenamefont {Katsnelson},\ and\ \citenamefont
  {Hussey}}]{2017s_ghannadzadeh_PdCrO2}%
  \BibitemOpen
  \bibfield  {author} {\bibinfo {author} {\bibfnamefont {S.}~\bibnamefont
  {Ghannadzadeh}}, \bibinfo {author} {\bibfnamefont {S.}~\bibnamefont
  {Licciardello}}, \bibinfo {author} {\bibfnamefont {S.}~\bibnamefont
  {Arsenijevi{\'c}}}, \bibinfo {author} {\bibfnamefont {P.}~\bibnamefont
  {Robinson}}, \bibinfo {author} {\bibfnamefont {H.}~\bibnamefont {Takatsu}},
  \bibinfo {author} {\bibfnamefont {M.}~\bibnamefont {Katsnelson}}, \ and\
  \bibinfo {author} {\bibfnamefont {N.}~\bibnamefont {Hussey}},\ }\href@noop {}
  {\bibfield  {journal} {\bibinfo  {journal} {Nature communications}\ }\textbf
  {\bibinfo {volume} {8}},\ \bibinfo {pages} {15001} (\bibinfo {year}
  {2017})}\BibitemShut {NoStop}%
\bibitem [{\citenamefont {Kadowaki}\ \emph {et~al.}(1995)\citenamefont
  {Kadowaki}, \citenamefont {Takei},\ and\ \citenamefont
  {Motoya}}]{1995_kadowaki_LiCrO2}%
  \BibitemOpen
  \bibfield  {author} {\bibinfo {author} {\bibfnamefont {H.}~\bibnamefont
  {Kadowaki}}, \bibinfo {author} {\bibfnamefont {H.}~\bibnamefont {Takei}}, \
  and\ \bibinfo {author} {\bibfnamefont {K.}~\bibnamefont {Motoya}},\
  }\href@noop {} {\bibfield  {journal} {\bibinfo  {journal} {Journal of
  Physics: Condensed Matter}\ }\textbf {\bibinfo {volume} {7}},\ \bibinfo
  {pages} {6869} (\bibinfo {year} {1995})}\BibitemShut {NoStop}%
\bibitem [{\citenamefont {Suzuki}\ \emph {et~al.}(2017)\citenamefont {Suzuki},
  \citenamefont {Koretsune}, \citenamefont {Ochi},\ and\ \citenamefont
  {Arita}}]{suzuki2017cluster}%
  \BibitemOpen
  \bibfield  {author} {\bibinfo {author} {\bibfnamefont {M.-T.}\ \bibnamefont
  {Suzuki}}, \bibinfo {author} {\bibfnamefont {T.}~\bibnamefont {Koretsune}},
  \bibinfo {author} {\bibfnamefont {M.}~\bibnamefont {Ochi}}, \ and\ \bibinfo
  {author} {\bibfnamefont {R.}~\bibnamefont {Arita}},\ }\href@noop {}
  {\bibfield  {journal} {\bibinfo  {journal} {Physical Review B}\ }\textbf
  {\bibinfo {volume} {95}},\ \bibinfo {pages} {094406} (\bibinfo {year}
  {2017})}\BibitemShut {NoStop}%
\bibitem [{\citenamefont {Nakatsuji}\ \emph {et~al.}(2015)\citenamefont
  {Nakatsuji}, \citenamefont {Kiyohara},\ and\ \citenamefont
  {Higo}}]{nakatsuji2015large}%
  \BibitemOpen
  \bibfield  {author} {\bibinfo {author} {\bibfnamefont {S.}~\bibnamefont
  {Nakatsuji}}, \bibinfo {author} {\bibfnamefont {N.}~\bibnamefont {Kiyohara}},
  \ and\ \bibinfo {author} {\bibfnamefont {T.}~\bibnamefont {Higo}},\
  }\href@noop {} {\bibfield  {journal} {\bibinfo  {journal} {Nature}\ }\textbf
  {\bibinfo {volume} {527}},\ \bibinfo {pages} {212} (\bibinfo {year}
  {2015})}\BibitemShut {NoStop}%
\bibitem [{\citenamefont {Toth}\ and\ \citenamefont
  {Lake}(2015)}]{2015_Toth_linear}%
  \BibitemOpen
  \bibfield  {author} {\bibinfo {author} {\bibfnamefont {S.}~\bibnamefont
  {Toth}}\ and\ \bibinfo {author} {\bibfnamefont {B.}~\bibnamefont {Lake}},\
  }\href@noop {} {\bibfield  {journal} {\bibinfo  {journal} {Journal of
  Physics: Condensed Matter}\ }\textbf {\bibinfo {volume} {27}},\ \bibinfo
  {pages} {166002} (\bibinfo {year} {2015})}\BibitemShut {NoStop}%
\bibitem [{\citenamefont {Mackenzie}(2017)}]{2017_mackenzie_delafossite}%
  \BibitemOpen
  \bibfield  {author} {\bibinfo {author} {\bibfnamefont {A.~P.}\ \bibnamefont
  {Mackenzie}},\ }\href@noop {} {\bibfield  {journal} {\bibinfo  {journal}
  {Reports on Progress in Physics}\ }\textbf {\bibinfo {volume} {80}},\
  \bibinfo {pages} {032501} (\bibinfo {year} {2017})}\BibitemShut {NoStop}%
\bibitem [{\citenamefont {Kim}\ \emph {et~al.}(2009)\citenamefont {Kim},
  \citenamefont {Choi},\ and\ \citenamefont {Min}}]{2009_KyooKim_PdCoO2}%
  \BibitemOpen
  \bibfield  {author} {\bibinfo {author} {\bibfnamefont {K.}~\bibnamefont
  {Kim}}, \bibinfo {author} {\bibfnamefont {H.~C.}\ \bibnamefont {Choi}}, \
  and\ \bibinfo {author} {\bibfnamefont {B.}~\bibnamefont {Min}},\ }\href@noop
  {} {\bibfield  {journal} {\bibinfo  {journal} {Physical Review B}\ }\textbf
  {\bibinfo {volume} {80}},\ \bibinfo {pages} {035116} (\bibinfo {year}
  {2009})}\BibitemShut {NoStop}%
\bibitem [{\citenamefont {Noh}\ \emph {et~al.}(2009)\citenamefont {Noh},
  \citenamefont {Jeong}, \citenamefont {Jeong}, \citenamefont {Cho},
  \citenamefont {Kim}, \citenamefont {Kim}, \citenamefont {Min},\ and\
  \citenamefont {Kim}}]{2009_HanJinNoh_PdCoO2}%
  \BibitemOpen
  \bibfield  {author} {\bibinfo {author} {\bibfnamefont {H.-J.}\ \bibnamefont
  {Noh}}, \bibinfo {author} {\bibfnamefont {J.}~\bibnamefont {Jeong}}, \bibinfo
  {author} {\bibfnamefont {J.}~\bibnamefont {Jeong}}, \bibinfo {author}
  {\bibfnamefont {E.-J.}\ \bibnamefont {Cho}}, \bibinfo {author} {\bibfnamefont
  {S.~B.}\ \bibnamefont {Kim}}, \bibinfo {author} {\bibfnamefont
  {K.}~\bibnamefont {Kim}}, \bibinfo {author} {\bibfnamefont {B.}~\bibnamefont
  {Min}}, \ and\ \bibinfo {author} {\bibfnamefont {H.-D.}\ \bibnamefont
  {Kim}},\ }\href@noop {} {\bibfield  {journal} {\bibinfo  {journal} {Physical
  review letters}\ }\textbf {\bibinfo {volume} {102}},\ \bibinfo {pages}
  {256404} (\bibinfo {year} {2009})}\BibitemShut {NoStop}%
\bibitem [{\citenamefont {Hicks}\ \emph {et~al.}(2015)\citenamefont {Hicks},
  \citenamefont {Gibbs}, \citenamefont {Zhao}, \citenamefont {Kushwaha},
  \citenamefont {Borrmann}, \citenamefont {Mackenzie}, \citenamefont {Takatsu},
  \citenamefont {Yonezawa}, \citenamefont {Maeno},\ and\ \citenamefont
  {Yelland}}]{2015_Hicks_PdCrO2}%
  \BibitemOpen
  \bibfield  {author} {\bibinfo {author} {\bibfnamefont {C.~W.}\ \bibnamefont
  {Hicks}}, \bibinfo {author} {\bibfnamefont {A.~S.}\ \bibnamefont {Gibbs}},
  \bibinfo {author} {\bibfnamefont {L.}~\bibnamefont {Zhao}}, \bibinfo {author}
  {\bibfnamefont {P.}~\bibnamefont {Kushwaha}}, \bibinfo {author}
  {\bibfnamefont {H.}~\bibnamefont {Borrmann}}, \bibinfo {author}
  {\bibfnamefont {A.~P.}\ \bibnamefont {Mackenzie}}, \bibinfo {author}
  {\bibfnamefont {H.}~\bibnamefont {Takatsu}}, \bibinfo {author} {\bibfnamefont
  {S.}~\bibnamefont {Yonezawa}}, \bibinfo {author} {\bibfnamefont
  {Y.}~\bibnamefont {Maeno}}, \ and\ \bibinfo {author} {\bibfnamefont {E.~A.}\
  \bibnamefont {Yelland}},\ }\href@noop {} {\bibfield  {journal} {\bibinfo
  {journal} {Physical Review B}\ }\textbf {\bibinfo {volume} {92}},\ \bibinfo
  {pages} {014425} (\bibinfo {year} {2015})}\BibitemShut {NoStop}%
\bibitem [{\citenamefont {Takatsu}\ \emph
  {et~al.}(2010{\natexlab{b}})\citenamefont {Takatsu}, \citenamefont
  {Yonezawa}, \citenamefont {Michioka}, \citenamefont {Yoshimura},\ and\
  \citenamefont {Maeno}}]{2010_Takatsu_anisotropy}%
  \BibitemOpen
  \bibfield  {author} {\bibinfo {author} {\bibfnamefont {H.}~\bibnamefont
  {Takatsu}}, \bibinfo {author} {\bibfnamefont {S.}~\bibnamefont {Yonezawa}},
  \bibinfo {author} {\bibfnamefont {C.}~\bibnamefont {Michioka}}, \bibinfo
  {author} {\bibfnamefont {K.}~\bibnamefont {Yoshimura}}, \ and\ \bibinfo
  {author} {\bibfnamefont {Y.}~\bibnamefont {Maeno}},\ }in\ \href@noop {}
  {\emph {\bibinfo {booktitle} {Journal of Physics: Conference Series}}},\
  Vol.\ \bibinfo {volume} {200}\ (\bibinfo {organization} {IOP Publishing},\
  \bibinfo {year} {2010})\ p.\ \bibinfo {pages} {012198}\BibitemShut {NoStop}%
\bibitem [{\citenamefont {Sunko}\ \emph {et~al.}(2018)\citenamefont {Sunko},
  \citenamefont {Mazzola}, \citenamefont {Kitamura}, \citenamefont {Khim},
  \citenamefont {Kushwaha}, \citenamefont {Clark}, \citenamefont {Watson},
  \citenamefont {Markovic}, \citenamefont {Biswas}, \citenamefont {Pourovskii}
  \emph {et~al.}}]{sunko2018probing}%
  \BibitemOpen
  \bibfield  {author} {\bibinfo {author} {\bibfnamefont {V.}~\bibnamefont
  {Sunko}}, \bibinfo {author} {\bibfnamefont {F.}~\bibnamefont {Mazzola}},
  \bibinfo {author} {\bibfnamefont {S.}~\bibnamefont {Kitamura}}, \bibinfo
  {author} {\bibfnamefont {S.}~\bibnamefont {Khim}}, \bibinfo {author}
  {\bibfnamefont {P.}~\bibnamefont {Kushwaha}}, \bibinfo {author}
  {\bibfnamefont {O.}~\bibnamefont {Clark}}, \bibinfo {author} {\bibfnamefont
  {M.}~\bibnamefont {Watson}}, \bibinfo {author} {\bibfnamefont
  {I.}~\bibnamefont {Markovic}}, \bibinfo {author} {\bibfnamefont
  {D.}~\bibnamefont {Biswas}}, \bibinfo {author} {\bibfnamefont
  {L.}~\bibnamefont {Pourovskii}},  \emph {et~al.},\ }\href@noop {} {\bibfield
  {journal} {\bibinfo  {journal} {arXiv preprint arXiv:1809.08972}\ } (\bibinfo
  {year} {2018})}\BibitemShut {NoStop}%
\bibitem [{\citenamefont {Ozaki}(2003)}]{2003_Ozaki_variationally}%
  \BibitemOpen
  \bibfield  {author} {\bibinfo {author} {\bibfnamefont {T.}~\bibnamefont
  {Ozaki}},\ }\href@noop {} {\bibfield  {journal} {\bibinfo  {journal}
  {Physical Review B}\ }\textbf {\bibinfo {volume} {67}},\ \bibinfo {pages}
  {155108} (\bibinfo {year} {2003})}\BibitemShut {NoStop}%
\bibitem [{\citenamefont {Ozaki}\ and\ \citenamefont
  {Kino}(2004)}]{2004_Ozaki_numerical}%
  \BibitemOpen
  \bibfield  {author} {\bibinfo {author} {\bibfnamefont {T.}~\bibnamefont
  {Ozaki}}\ and\ \bibinfo {author} {\bibfnamefont {H.}~\bibnamefont {Kino}},\
  }\href@noop {} {\bibfield  {journal} {\bibinfo  {journal} {Physical Review
  B}\ }\textbf {\bibinfo {volume} {69}},\ \bibinfo {pages} {195113} (\bibinfo
  {year} {2004})}\BibitemShut {NoStop}%
\bibitem [{\citenamefont {Han}\ \emph {et~al.}(2006)\citenamefont {Han},
  \citenamefont {Ozaki},\ and\ \citenamefont {Yu}}]{2006_han_n}%
  \BibitemOpen
  \bibfield  {author} {\bibinfo {author} {\bibfnamefont {M.~J.}\ \bibnamefont
  {Han}}, \bibinfo {author} {\bibfnamefont {T.}~\bibnamefont {Ozaki}}, \ and\
  \bibinfo {author} {\bibfnamefont {J.}~\bibnamefont {Yu}},\ }\href@noop {}
  {\bibfield  {journal} {\bibinfo  {journal} {Physical Review B}\ }\textbf
  {\bibinfo {volume} {73}},\ \bibinfo {pages} {045110} (\bibinfo {year}
  {2006})}\BibitemShut {NoStop}%
\bibitem [{\citenamefont {Kurz}\ \emph {et~al.}(2004)\citenamefont {Kurz},
  \citenamefont {F{\"o}rster}, \citenamefont {Nordstr{\"o}m}, \citenamefont
  {Bihlmayer},\ and\ \citenamefont {Bl{\"u}gel}}]{kurz2004ab}%
  \BibitemOpen
  \bibfield  {author} {\bibinfo {author} {\bibfnamefont {P.}~\bibnamefont
  {Kurz}}, \bibinfo {author} {\bibfnamefont {F.}~\bibnamefont {F{\"o}rster}},
  \bibinfo {author} {\bibfnamefont {L.}~\bibnamefont {Nordstr{\"o}m}}, \bibinfo
  {author} {\bibfnamefont {G.}~\bibnamefont {Bihlmayer}}, \ and\ \bibinfo
  {author} {\bibfnamefont {S.}~\bibnamefont {Bl{\"u}gel}},\ }\href@noop {}
  {\bibfield  {journal} {\bibinfo  {journal} {Physical Review B}\ }\textbf
  {\bibinfo {volume} {69}},\ \bibinfo {pages} {024415} (\bibinfo {year}
  {2004})}\BibitemShut {NoStop}%
\bibitem [{\citenamefont {Kresse}\ and\ \citenamefont
  {Furthm{\"u}ller}(1996)}]{kresse1996efficient}%
  \BibitemOpen
  \bibfield  {author} {\bibinfo {author} {\bibfnamefont {G.}~\bibnamefont
  {Kresse}}\ and\ \bibinfo {author} {\bibfnamefont {J.}~\bibnamefont
  {Furthm{\"u}ller}},\ }\href@noop {} {\bibfield  {journal} {\bibinfo
  {journal} {Physical review B}\ }\textbf {\bibinfo {volume} {54}},\ \bibinfo
  {pages} {11169} (\bibinfo {year} {1996})}\BibitemShut {NoStop}%
\bibitem [{\citenamefont {Ceperley}\ and\ \citenamefont
  {Alder}(1980)}]{ceperley1980ground}%
  \BibitemOpen
  \bibfield  {author} {\bibinfo {author} {\bibfnamefont {D.~M.}\ \bibnamefont
  {Ceperley}}\ and\ \bibinfo {author} {\bibfnamefont {B.}~\bibnamefont
  {Alder}},\ }\href@noop {} {\bibfield  {journal} {\bibinfo  {journal}
  {Physical Review Letters}\ }\textbf {\bibinfo {volume} {45}},\ \bibinfo
  {pages} {566} (\bibinfo {year} {1980})}\BibitemShut {NoStop}%
\bibitem [{\citenamefont {Liechtenstein}\ \emph {et~al.}(1995)\citenamefont
  {Liechtenstein}, \citenamefont {Anisimov},\ and\ \citenamefont
  {Zaanen}}]{liechtenstein1995density}%
  \BibitemOpen
  \bibfield  {author} {\bibinfo {author} {\bibfnamefont {A.}~\bibnamefont
  {Liechtenstein}}, \bibinfo {author} {\bibfnamefont {V.}~\bibnamefont
  {Anisimov}}, \ and\ \bibinfo {author} {\bibfnamefont {J.}~\bibnamefont
  {Zaanen}},\ }\href@noop {} {\bibfield  {journal} {\bibinfo  {journal}
  {Physical Review B}\ }\textbf {\bibinfo {volume} {52}},\ \bibinfo {pages}
  {R5467} (\bibinfo {year} {1995})}\BibitemShut {NoStop}%
\bibitem [{\citenamefont {T{\'o}th}\ \emph {et~al.}(2016)\citenamefont
  {T{\'o}th}, \citenamefont {Wehinger}, \citenamefont {Rolfs}, \citenamefont
  {Birol}, \citenamefont {Stuhr}, \citenamefont {Takatsu}, \citenamefont
  {Kimura}, \citenamefont {Kimura}, \citenamefont {R{\o}nnow},\ and\
  \citenamefont {R{\"u}egg}}]{toth2016electromagnon}%
  \BibitemOpen
  \bibfield  {author} {\bibinfo {author} {\bibfnamefont {S.}~\bibnamefont
  {T{\'o}th}}, \bibinfo {author} {\bibfnamefont {B.}~\bibnamefont {Wehinger}},
  \bibinfo {author} {\bibfnamefont {K.}~\bibnamefont {Rolfs}}, \bibinfo
  {author} {\bibfnamefont {T.}~\bibnamefont {Birol}}, \bibinfo {author}
  {\bibfnamefont {U.}~\bibnamefont {Stuhr}}, \bibinfo {author} {\bibfnamefont
  {H.}~\bibnamefont {Takatsu}}, \bibinfo {author} {\bibfnamefont
  {K.}~\bibnamefont {Kimura}}, \bibinfo {author} {\bibfnamefont
  {T.}~\bibnamefont {Kimura}}, \bibinfo {author} {\bibfnamefont {H.~M.}\
  \bibnamefont {R{\o}nnow}}, \ and\ \bibinfo {author} {\bibfnamefont
  {C.}~\bibnamefont {R{\"u}egg}},\ }\href@noop {} {\bibfield  {journal}
  {\bibinfo  {journal} {Nature communications}\ }\textbf {\bibinfo {volume}
  {7}},\ \bibinfo {pages} {13547} (\bibinfo {year} {2016})}\BibitemShut
  {NoStop}%
\bibitem [{\citenamefont {Jain}\ \emph {et~al.}(2013)\citenamefont {Jain},
  \citenamefont {Ong}, \citenamefont {Hautier}, \citenamefont {Chen},
  \citenamefont {Richards}, \citenamefont {Dacek}, \citenamefont {Cholia},
  \citenamefont {Gunter}, \citenamefont {Skinner}, \citenamefont {Ceder} \emph
  {et~al.}}]{jain2013commentary}%
  \BibitemOpen
  \bibfield  {author} {\bibinfo {author} {\bibfnamefont {A.}~\bibnamefont
  {Jain}}, \bibinfo {author} {\bibfnamefont {S.~P.}\ \bibnamefont {Ong}},
  \bibinfo {author} {\bibfnamefont {G.}~\bibnamefont {Hautier}}, \bibinfo
  {author} {\bibfnamefont {W.}~\bibnamefont {Chen}}, \bibinfo {author}
  {\bibfnamefont {W.~D.}\ \bibnamefont {Richards}}, \bibinfo {author}
  {\bibfnamefont {S.}~\bibnamefont {Dacek}}, \bibinfo {author} {\bibfnamefont
  {S.}~\bibnamefont {Cholia}}, \bibinfo {author} {\bibfnamefont
  {D.}~\bibnamefont {Gunter}}, \bibinfo {author} {\bibfnamefont
  {D.}~\bibnamefont {Skinner}}, \bibinfo {author} {\bibfnamefont
  {G.}~\bibnamefont {Ceder}},  \emph {et~al.},\ }\href@noop {} {\bibfield
  {journal} {\bibinfo  {journal} {Apl Materials}\ }\textbf {\bibinfo {volume}
  {1}},\ \bibinfo {pages} {011002} (\bibinfo {year} {2013})}\BibitemShut
  {NoStop}%
\bibitem [{\citenamefont {Moriya}(1960)}]{moriya1960anisotropic}%
  \BibitemOpen
  \bibfield  {author} {\bibinfo {author} {\bibfnamefont {T.}~\bibnamefont
  {Moriya}},\ }\href@noop {} {\bibfield  {journal} {\bibinfo  {journal}
  {Physical Review}\ }\textbf {\bibinfo {volume} {120}},\ \bibinfo {pages} {91}
  (\bibinfo {year} {1960})}\BibitemShut {NoStop}%
\bibitem [{\citenamefont {Kushwaha}\ \emph {et~al.}(2015)\citenamefont
  {Kushwaha}, \citenamefont {Sunko}, \citenamefont {Moll}, \citenamefont
  {Bawden}, \citenamefont {Riley}, \citenamefont {Nandi}, \citenamefont
  {Rosner}, \citenamefont {Schmidt}, \citenamefont {Arnold}, \citenamefont
  {Hassinger} \emph {et~al.}}]{kushwaha2015nearly}%
  \BibitemOpen
  \bibfield  {author} {\bibinfo {author} {\bibfnamefont {P.}~\bibnamefont
  {Kushwaha}}, \bibinfo {author} {\bibfnamefont {V.}~\bibnamefont {Sunko}},
  \bibinfo {author} {\bibfnamefont {P.~J.}\ \bibnamefont {Moll}}, \bibinfo
  {author} {\bibfnamefont {L.}~\bibnamefont {Bawden}}, \bibinfo {author}
  {\bibfnamefont {J.~M.}\ \bibnamefont {Riley}}, \bibinfo {author}
  {\bibfnamefont {N.}~\bibnamefont {Nandi}}, \bibinfo {author} {\bibfnamefont
  {H.}~\bibnamefont {Rosner}}, \bibinfo {author} {\bibfnamefont {M.~P.}\
  \bibnamefont {Schmidt}}, \bibinfo {author} {\bibfnamefont {F.}~\bibnamefont
  {Arnold}}, \bibinfo {author} {\bibfnamefont {E.}~\bibnamefont {Hassinger}},
  \emph {et~al.},\ }\href@noop {} {\bibfield  {journal} {\bibinfo  {journal}
  {Science advances}\ }\textbf {\bibinfo {volume} {1}},\ \bibinfo {pages}
  {e1500692} (\bibinfo {year} {2015})}\BibitemShut {NoStop}%
\bibitem [{\citenamefont {Wang}\ \emph {et~al.}(2006)\citenamefont {Wang},
  \citenamefont {Yates}, \citenamefont {Souza},\ and\ \citenamefont
  {Vanderbilt}}]{wang2006ab}%
  \BibitemOpen
  \bibfield  {author} {\bibinfo {author} {\bibfnamefont {X.}~\bibnamefont
  {Wang}}, \bibinfo {author} {\bibfnamefont {J.~R.}\ \bibnamefont {Yates}},
  \bibinfo {author} {\bibfnamefont {I.}~\bibnamefont {Souza}}, \ and\ \bibinfo
  {author} {\bibfnamefont {D.}~\bibnamefont {Vanderbilt}},\ }\href@noop {}
  {\bibfield  {journal} {\bibinfo  {journal} {Physical Review B}\ }\textbf
  {\bibinfo {volume} {74}},\ \bibinfo {pages} {195118} (\bibinfo {year}
  {2006})}\BibitemShut {NoStop}%
\end{thebibliography}
%

\end{document}